\documentclass[prd,nofootinbib,reprint,twocolumn,showpacs,superscriptaddress,floatfix]{revtex4-1}
\pdfoutput=1 
\usepackage{bm, color} 
\usepackage{amssymb,amsfonts,slashed,amsthm,amsmath, 
graphicx,xcolor, soul, empheq, cancel}

\usepackage{hyperref}

\usepackage[caption=false]{subfig}

\begin{document}

\newcommand{\vev}[1]{ \left\langle {#1} \right\rangle }
\newcommand{\bra}[1]{ \langle {#1} | }
\newcommand{\ket}[1]{ | {#1} \rangle }
\newcommand{\eV}{ \ {\rm eV} }
\newcommand{\KeV}{ \ {\rm keV} }
\newcommand{\MeV}{\  {\rm MeV} }
\newcommand{\GeV}{\  {\rm GeV} }
\newcommand{\TeV}{\  {\rm TeV} }
\newcommand{\1}{\mbox{1}\hspace{-0.25em}\mbox{l}}
\newcommand{\Red}[1]{{\color{red} {#1}}}

\newcommand{\lmk}{\left(}  
\newcommand{\rmk}{\right)}
\newcommand{\lkk}{\left[}  
\newcommand{\rkk}{\right]}
\newcommand{\lhk}{\left \{ }  
\newcommand{\rhk}{\right \} }
\newcommand{\del}{\partial}  
\newcommand{\la}{\left\langle} 
\newcommand{\ra}{\right\rangle}
\newcommand{\half}{\frac{1}{2}}

\newcommand{\bea}{\begin{array}}
\newcommand{\eea}{\end{array}}
\newcommand{\beq}{\begin{eqnarray}}
\newcommand{\eeq}{\end{eqnarray}}
\newcommand{\eq}[1]{Eq.~(\ref{#1})}

\newcommand{\dd}{\mathrm{d}}
\newcommand{\Mpl}{M_{\rm Pl}}
\newcommand{\mg}{m_{3/2}}
\newcommand{\abs}[1]{\left\vert {#1} \right\vert}
\newcommand{\mphi}{m_{\phi}}
\newcommand{\Hz}{\ {\rm Hz}}
\newcommand{\for}{\quad \text{for }}
\newcommand{\Min}{\text{Min}}
\newcommand{\Max}{\text{Max}}
\newcommand{\Kahler}{K\"{a}hler }
\newcommand{\cphi}{\varphi}
\newcommand{\Tr}{\text{Tr}}
\newcommand{\diag}{{\rm diag}}

\newcommand{\SUf}{SU(3)_{\rm f}}
\newcommand{\Upq}{U(1)_{\rm PQ}}
\newcommand{\Zpq}{Z^{\rm PQ}_3}
\newcommand{\Cpq}{C_{\rm PQ}}
\newcommand{\ubar}{u^c}
\newcommand{\dbar}{d^c}
\newcommand{\ebar}{e^c}
\newcommand{\nubar}{\nu^c}
\newcommand{\Ndw}{N_{\rm DW}}
\newcommand{\Fpq}{F_{\rm PQ}}
\newcommand{\fpq}{v_{\rm PQ}}
\newcommand{\Br}{{\rm Br}}
\newcommand{\Lag}{\mathcal{L}}
\newcommand{\Lqcd}{\Lambda_{\rm QCD}}

\newcommand{\ji}{j_{\rm inf}} 
\newcommand{\jb}{j_{B-L}} 
\newcommand{\M}{M} 
\newcommand{\im}{{\rm Im} }
\newcommand{\re}{{\rm Re} }

\def\lrf#1#2{ \left(\frac{#1}{#2}\right)}
\def\lrfp#1#2#3{ \left(\frac{#1}{#2} \right)^{#3}}
\def\lrp#1#2{\left( #1 \right)^{#2}}
\def\REF#1{Ref.~\cite{#1}}
\def\SEC#1{Sec.~\ref{#1}}
\def\FIG#1{Fig.~\ref{#1}}
\def\EQ#1{Eq.~(\ref{#1})}
\def\EQS#1{Eqs.~(\ref{#1})}
\def\TEV#1{10^{#1}{\rm\,TeV}}
\def\GEV#1{10^{#1}{\rm\,GeV}}
\def\MEV#1{10^{#1}{\rm\,MeV}}
\def\KEV#1{10^{#1}{\rm\,keV}}
\def\blue#1{\textcolor{blue}{#1}}
\def\red#1{\textcolor{blue}{#1}}

\newcommand{\eff}{\Delta N_{\rm eff}}
\newcommand{\neff}{\Delta N_{\rm eff}}
\newcommand{\cc}{\Omega_\Lambda}
\newcommand{\Mpc}{\ {\rm Mpc}}
\newcommand{\Msolar}{M_\odot}

\newcommand{\er}[1]{Eq.~\eqref{#1}}
\newcommand{\fr}{\frac}
\newcommand{\der}{\partial}
\newcommand{\bb}{\mathbb}

\newcommand{\RYcom}[1]{{\bf\color[HTML]{0000CC} (RY: #1)}}
\newcommand{\RY}[1]{{\bf\color[HTML]{0000CC}#1}}

\newcommand{\MScom}[1]{{\bf\color[HTML]{FFA500} (MS: #1)}}


\title{
Lazarides-Shafi axion models as Dijkgraaf-Witten theories
}
\author{Motoo Suzuki$^{1,2,3}$ and Ryo Yokokura$^{4}$
\\*[10pt]
$^1${\it \normalsize SISSA International School for Advanced Studies, \\
Via Bonomea 265, 34136, Trieste, Italy} \\
$^2${\it \normalsize INFN, Sezione di Trieste, Via Valerio 2, 34127, Italy}\\
$^3${\it \normalsize IFPU, Via Beirut 2, 34014 Trieste, Italy}\\
$^4${\it \normalsize Department of Physics \& Research and Education Center for Natural Sciences, Keio University, Yokohama 223-8521, Japan}
}

\begin{abstract}
Axion models often face the domain wall problem, which threatens the standard big-bang cosmology.
The Lazarides–Shafi mechanism attempts to resolve this by identifying degenerate vacua through a continuous gauge symmetry.
We formulate a topological quantum field theory to isolate the essential structure of the mechanism and analyze its generalized symmetry structure, including higher-form symmetries and higher-group.
This framework yields a master formula for computing the domain wall number and clarifies the higher-form symmetry conditions required for complete vacuum identification in a model independent way.
Moreover, while a domain-wall-number-one scenario eliminates all higher-form global symmetries, 
the theory nevertheless exhibits a nontrivial four-group structure and realizes a symmetry-protected topological (SPT) phase.
\end{abstract}

\maketitle


\renewcommand{\thefootnote}{$*$\arabic{footnote}}%
\section{Introduction}
\label{introduction}
The axion is a hypothetical pseudo-Nambu–Goldstone boson introduced to solve the strong CP problem via the Peccei–Quinn mechanism~\cite{Peccei:1977hh, Peccei:1977ur, Weinberg:1977ma, Wilczek:1977pj}.
Axion models often suffer from the stable domain wall problem, which can spoil standard big-bang cosmology~\cite{Sikivie:1982qv}. The Lazarides–Shafi mechanism~\cite{Lazarides:1982tw} proposes to resolve this issue by identifying degenerate vacua through a continuous gauge symmetry, thereby avoiding stable walls.%
\footnote{Identifying the vacua alone does not suffice to confirm the domain wall problem is resolved; simulations are also needed to capture the string-wall dynamics~\cite{Hiramatsu:2019tua,Hiramatsu:2020zlp}.}
Yet, recent analyses~\cite{Lu:2023ayc} have shown that in some models the vacuum identification is incomplete due to subtle global-structure issues, and the domain wall problem persists.

To clarify the conditions under which the mechanism succeeds, we construct a topological quantum field theory (TQFT) that isolates its essential structure and analyze the generalized global symmetries~\cite{Kapustin:2014gua,Gaiotto:2014kfa} in the framework (see e.g.~\cite{Cordova:2022ruw, McGreevy:2022oyu, Gomes:2023ahz, Schafer-Nameki:2023jdn, Brennan:2023mmt, Luo:2023ive, Shao:2023gho, Costa:2024wks, Iqbal:2024pee, Davighi:2025iyk} for reviews). 
The resulting TQFT belongs to the class of Dijkgraaf–Witten theories~\cite{Dijkgraaf:1989pz}.
This approach yields a master formula to compute the domain wall number of the axion in a systematic way, and also puts constraints on higher-form symmetries for successful Lazarides--Shafi mechanism. In particular, successful models require breaking of a certain one-form symmetry. 
These conditions provide a general criterion for diagnosing the domain wall problem in axion models, particularly those that also address the axion quality problem~\cite{Georgi:1981pu,Kamionkowski:1992mf,Holman:1992us,Kallosh:1995hi,Barr:1992qq,Ghigna:1992iv,Dine:1992vx}.
We also discuss that the theory is characterized 
by a four-group structure~\cite{Hidaka:2021mml,Hidaka:2021kkf} and in the symmetry-protected topological (SPT) phase, giving the physical interpretation of them.%
\footnote{The higher-group structure in the axion-photon system has been investigated in~\cite{Hidaka:2020iaz,Hidaka:2020izy,Brennan:2020ehu,Hidaka:2021mml,Hidaka:2021kkf}.}

\section{TQFT describing the vacuum identification in axion models}

In general, axion models possess an anomaly-free global zero-form discrete $\mathbb{Z}_{N}$ symmetry embedded in the $U(1)$ Peccei–Quinn symmetry.
The Lazarides–Shafi mechanism embeds this $\mathbb{Z}_N$ symmetry into the center of a continuous gauge symmetry such as $SU(N)$ or $U(1)$.%
\footnote{There is another class of the Lazarides--Shafi mechanism,
where a subgroup that is not a center identifies the vacua
such as $U(1)$ subgroup of $SU(2)$~\cite{Sato:2018nqy,Chatterjee:2019rch}.
In this mechanism, the axionic domain walls can shrink by producing 
Alice strings~\cite{Schwarz:1982ec}.}
In the infrared, the continuous gauge symmetry does not manifest due to Higgsing or confinement, but a zero-form $\mathbb{Z}_N$ discrete gauge symmetry may remain, under which the $N$ degenerate axion vacua are identified.

We propose a 4d TQFT action describing a system of $N_1$ degenerate vacua 
in the low-energy limit, among which a $\mathbb{Z}_{N_2}$ center symmetry identifies certain subsets:
\begin{align}
\label{eq:tqft_1}
    S=\frac{1}{2\pi}
    \int N_1\phi dC
    +N_2 B\wedge dA+{\rm lcm}(N_1,N_2)\,M C\wedge A\ .
\end{align}
Here, $\phi$ is the $2\pi$-periodic zero-form field representing the axion, $A$ is a one-form $U(1)$ gauge field, $B$ is a two-form $U(1)$ gauge field, and $C$ is a three-form $U(1)$ gauge field arising from the QCD topological term 
$\frac{N_1}{8\pi^2}\phi \, \mathrm{Tr}(G \wedge G)$
with the gluon field strength $G$.
The numbers 
$N_1$, $N_2$, and $M$ are positive integers unless otherwise noted.
${\rm lcm}(N_1,N_2)$ denotes the least common multiple of $N_1$ and $N_2$; it is introduced for convenience, and omitting it does not affect the generality of the coefficients. 
We have neglected the kinetic terms of the dynamical fields,
since we are in the low-energy limit and 
no propagating degree of freedom remains apart from the axion. 
This is a TQFT in the sense that the action in Eq.~\eqref{eq:tqft_1} is metric-independent.

The first term in Eq.~\eqref{eq:tqft_1} corresponds to the 
axion–topological ($\theta$) coupling in 
QCD, where $C$ effectively describes the 
Chern-Simons term~\cite{Aurilia:1980jz,DiVecchia:1980yfw,Dvali:2005an}.
This term can also be derived from the axion potential with $N_1$ degenerate vacua~\cite{Hidaka:2019mfm}.
The second term represents the $\mathbb{Z}_{N_2}$ gauge sector
by restricting the $U(1)$ gauge field $A$ to the $\mathbb{Z}_{N_2}$
gauge field.%
\footnote{In the case of $SU(N)$, the $U(1)$ gauge field $A$ is introduced by extending the theory to $U(N) = [SU(N)\times U(1)]/\mathbb{Z}_N$, and the second term in Eq.~\eqref{eq:tqft_1} reduces the $U(1)$ gauge symmetry to $\mathbb{Z}_N$.}
The last term links the two sectors
by modifying $d \phi \wedge C$ with a Stueckelberg coupling
 $d\phi \to d\phi + \ell A$ with some integer $\ell$, realizing the vacuum identification, as we discuss below.
The identification of vacua may be incomplete depending on the parameters $N_1$, $N_2$, and $M$.
We will derive the condition under which the vacua are completely identified by the center symmetry in the next section.

We remark that this TQFT is model independent in the sense that 
the action can be determined by the symmetries of 
the axion and the gauge field, rather than the details of models. 
Further, this 
TQFT action represents 
a 
4d generalization of the 2d Dijkgraaf–Witten theory~\cite{Dijkgraaf:1989pz} with $\mathbb{Z}_{N_1}\times \mathbb{Z}_{N_2}$ symmetries.
The system can be analyzed in parallel with the discussion in Appendix A of~\cite{Gaiotto:2014kfa}.

The action is invariant under the gauge transformations
\begin{equation}
\begin{aligned}
C &\to C + df_2,\quad
A \to A + df_0, \\
\phi &\to \phi - \frac{P}{K}N_2 f_0,\quad
B \to B + \frac{P}{K}N_1 f_2\ .
 \label{eq:gauge}
\end{aligned}
\end{equation}
and also under
\begin{align}
\phi\to \phi+2\pi\ ,\quad B\to B+df_1\ .
\end{align}
Here, $f_i$ denotes an $i$-form gauge parameter, and we have defined
\begin{align}
&P=\frac{M}{{\rm gcd}(N_1,N_2,M)}\ ,\quad
K=\frac{{\rm gcd}(N_1,N_2)}{{\rm gcd}(N_1,N_2,M)}\ ,\\
& {\rm gcd}(P,K)=1\ .
\end{align}

For $M = 0$, we identify the genuine (i.e.~gauge-invariant) point, line, surface, and volume operators, respectively,
\begin{align}
e^{i\phi}, \quad e^{i\oint A}, \quad e^{i\oint B}, \quad e^{i\oint C}\ .
\end{align}
These operators serve as the symmetry generators of the following global symmetries because of the equations of motion~\cite{Kapustin:2014gua}:
\begin{align}
\mathbb{Z}_{N_1}^{(3)} \times \mathbb{Z}_{N_2}^{(2)} \times \mathbb{Z}_{N_2}^{(1)} \times \mathbb{Z}_{N_1}^{(0)}\ ,
\end{align}
with the superscript $(i)$ indicating the $i$-form symmetry.
The corresponding charged operators are given by
\begin{align}
e^{i l_1 \oint C}, \quad e^{i l_2 \oint B}, \quad e^{i l_3 \oint A}, \quad e^{i l_4 \phi}\ ,
\end{align}
where $l_i$ are integers.

For $M \neq 0$, additional gauge transformations appear in \eqref{eq:gauge}, which reduce the remaining global symmetries.
The topological operators are now 
\begin{align}
    U_1(V)&= e^{i\oint_V C}\label{eq:U1}\\
    U_2(\gamma)&=e^{i\oint_\gamma A}\label{eq:U2}\\
   W_1(L_1) &=  e^{i\phi(P)} e^{i\frac{PN_2}{K}\int_{L_1}A}e^{-i\phi(P')},\label{eq:w1}
   \\
     W_2(D_3)&=e^{i\oint B} e^{-i\frac{P N_1}{K}\int_{D_3} C}\ ,\label{eq:w2}
\end{align}
where $L_1$ is a open line starting at the point $P$ and ending at $P'$, and $D_3$ is a three-dimensional 
subspace 
with 
a boundary.
The symmetries of this system can be analyzed by identifying genuine point, line, surface, and volume operators, yielding
\begin{align}
\mathbb{Z}_{N_1/K}^{(0)} \times \mathbb{Z}_{N_2/K}^{(1)} \times \mathbb{Z}_{N_2/K}^{(2)} \times \mathbb{Z}_{N_1/K}^{(3)}\ .
\end{align}
See the appendix~\ref{app:high_sym} for a more detailed discussion.
The reduction of the zero-form symmetry from $\mathbb{Z}_{N_1}^{(0)}$ to $\mathbb{Z}_{N_1/K}^{(0)}$ indicates that our TQFT action describes the vacuum identification realized by the Lazarides–Shafi mechanism.

The symmetries can be derived by integrating out certain fields, reducing the action to simpler forms.  
Let us first integrate out $B$ in the original TQFT, yielding
\begin{align}
    A = \frac{d\chi}{N_2}, \quad \chi \sim \chi + 2\pi,
\end{align}
and the TQFT action becomes
\begin{align}
\label{eq:int_B}
    \frac{1}{2\pi} \int \Big(N_1 \phi - {\rm lcm}(N_1, N_2) \frac{M}{N_2} \chi \Big) dC \ .
\end{align}
The gauge-invariant axion is then defined as a linear combination of $\phi$ and $\chi$.  
The $N_1$ degenerate vacua are reduced to
\begin{align}
{\rm gcd}\Big(N_1, {\rm lcm}(N_1,N_2) \frac{M}{N_2}\Big) =\frac{N_1}{K},
\end{align}
indicating that the global zero-form symmetry is $\mathbb{Z}^{(0)}_{N_1/K}$.  
The three-form global symmetry follows from the conditions
\begin{align}
    N_1 dC = 0, \quad {\rm lcm}(N_1,N_2) \frac{M}{N_2} dC = 0,
\end{align}
which imply a $\mathbb{Z}^{(3)}_{N_1/K}$ symmetry.

In a similar way, one can integrate out $\phi$ instead.
The action then reduces to a BF theory with $C = N_1 d\tilde B$, where $\tilde B$ is a $U(1)$ two-form gauge field:
\begin{align}
S = \frac{1}{2\pi} \int \big(N_2 B - \mathrm{lcm}(N_1, N_2)\frac{M}{N_1} \tilde B \big) \wedge dA\ .
\end{align}
This implies that the theory has the following global symmetries:
\begin{align}
 \mathbb{Z}_{N_2/K}^{(1)}\times \mathbb{Z}_{N_2/K}^{(2)}\ .
\end{align}

When $N_1/K=N_2/K=1$, there 
exist 
no global higher-form symmetries. 
Nonetheless, we observe non-trivial correlations between topological operators via 
defect crossing, i.e. $\langle W_1(l) W_2(D_3)\rangle=e^{-i2\pi\frac{P}{K}\int \delta(l)\wedge\delta(D_3)}$,
where $\delta (\Sigma)$ is a delta function $(4-q)$-form 
on a $q$-dimensional subspace $\Sigma$,
which satisfies $\int_{\Sigma} \omega_q = \int \omega_q \wedge \delta (\Sigma)$ for a $q$-form $\omega_q$.
Formally, these correlations correspond to the SPT phase%
\footnote{In the SPT phase, when the space-time manifold has a boundary, the bulk topological action is not gauge invariant under the background gauge field transformations.
Its gauge variation localizes on the boundary and is compensated by the ’t Hooft anomaly of the boundary degrees of freedom.
This is the anomaly-inflow mechanism through which the combined bulk–boundary system is fully gauge invariant.
}
associated with a four-group structure.
See Appendix~\ref{app:four_group} for a more detailed discussion about the four-group.

\section{Domain Wall Number and Higher-Form Symmetry Conditions}
The TQFT action in Eq.~\eqref{eq:tqft_1} provides a unified framework for computing the domain wall number and for identifying the higher-form symmetry conditions for complete vacuum identification.\\

\noindent
{\bf Master formula for the domain wall number:} 
The domain wall number of the axion, $N_{\rm DW}$, is determined by the remaining global discrete zero-form symmetry $\mathbb{Z}_{N_1/K}^{(0)}$:
\begin{align}
    N_{\rm DW} = \frac{N_1}{K}
    = N_1\, \frac{{\rm gcd}(N_1, N_2, M)}{{\rm gcd}(N_1, N_2)}\ .
    \label{eq:dom_num}
\end{align}

In a given model, $N_1$, $N_2$, and $M$ are determined as follows: $N_2$ denotes the center symmetry of the gauge group $G$ that participates in the vacuum identification; $N_1$ is obtained from the anomaly coefficient of the PQ symmetry with QCD, $\mathcal{A}(U(1)_{\rm PQ}-SU(3)_c^2)$; and $M$ is deduced by computing the (gauge-equivalent) shift of $\theta$ under the center symmetry transformation of $G$.  
While the master formula has been derived in previous literature~\cite{Lu:2023ayc,Azatov:2025mep}, our current version generalizes these earlier results.

Consider a theory containing Weyl fermions $\psi_j$ in irreducible representations of the gauge groups, with integer PQ charges $p_j$ whose greatest common divisor is $1$, transforming under the PQ symmetry as
\begin{align}
    \psi_j \;\to\; \psi_j\, e^{i \alpha_{\rm PQ} p_j}.
\end{align}
where $\alpha_{\rm PQ}$ is a real parameter. Under this transformation, the QCD $\theta$ term shifts according to
\begin{align}
    \int \frac{1}{8\pi^2} \theta \, {\rm tr}(G \wedge G)
    \;\to\;
    \int \frac{1}{8\pi^2} (\theta + \alpha_{\rm PQ}\, k_G)\, {\rm tr}(G \wedge G)\ .
\end{align} 
The maximal anomaly-free subgroup of $U(1)_{\rm PQ}$ is therefore $\mathbb{Z}_{k_G}$, giving
\begin{align}
    N_1 = k_G= \sum_i p_i \, {\rm dim}(r_i) \, 2 I(s_i) \ .
\end{align}
In this expression, ${\rm dim}(r_i)$ counts the dimension of the fermion representation under gauge groups other than $SU(3)_c$, while $I(s_i)$ is the Dynkin index for the $SU(3)_c$ representation, with the fundamental normalized to $1/2$.

The parameter $M$ is determined by the action of the center symmetry of the gauge group $G$, under which the fermions transform as
\begin{align}
\label{eq:center_transform}
    \psi_j \;\to\; \psi_j \, \exp\!\Big(\frac{i 2 \pi c_j}{N_2}\Big)\ ,
\end{align}
with $c_j$ denoting the center charge of $\psi_j$. This transformation induces a gauge-equivalent shift of the $\theta$ term,
\begin{align}
\int \frac{1}{8\pi^2} \theta \, {\rm tr}(G \wedge G)
&\;\to\;
\int \frac{1}{8\pi^2} (\theta + 2\pi k_c)\, {\rm tr}(G \wedge G),
\end{align}
with
\begin{align}
k_c &=
\sum_i \frac{c_i}{N_2} \, {\rm dim}(r_i) \, 2 I(s_i) \ .
\end{align}
This corresponds to the transformation $\chi \to \chi + 2\pi$ in Eq.~\eqref{eq:int_B}, and hence we find
\begin{align}
    {\rm lcm}(N_1, N_2) \frac{M}{N_2} = k_c \ .
\end{align}

Thus, by evaluating the shifts of $\theta$ under both the PQ and center symmetry transformations, 
one can systematically determine $N_1$ and $M$, and thereby compute the domain wall number with~\eqref{eq:dom_num}.\\

\noindent
{\bf Higher-form symmetries for $N_{\rm DW}=1$:} 
The successful implementation of the Lazarides--Shafi mechanism, corresponding to $N_{\rm DW}=1$, requires $N_1/K=1$, 
which is equivalent to
\begin{align}
    {\rm gcd}(N_1,M)=1 \quad \text{and} \quad {\rm gcd}(N_1,N_2)=N_1\ .
\end{align}
The remaining global symmetries are then
\begin{align}
    \mathbb{Z}_{N_2/N_1}^{(1)} \times \mathbb{Z}_{N_2/N_1}^{(2)}\ .
\end{align}
In the standard Lazarides--Shafi setup, $N_1=N_2$ is imposed, thereby eliminating all higher-form symmetries in the IR.  
The more general condition above, however, reveals a broader possibility: a smaller subgroup of the center symmetry can suffice to achieve vacuum identification.

This observation has a concrete implication for model building.  
One can examine the pattern of Higgsing or confinement from $G$ to a subgroup $H$ and analyze the one-form center symmetry in the resulting IR theory.  
If a residual one-form center symmetry via $G$ remains, it indicates $N_{\rm DW}>1$, implying that the theory suffers from the domain wall problem.  
In the next section, we show that this criterion allows the presence of a domain wall problem to be identified without explicitly computing the domain wall number.

\section{Applications to Explicit Models}\label{sec:model}
We consider two illustrative classes of models.  

First, we discuss the case $G=SU(N)$~\cite{Ardu:2020qmo}, in which an $SU(N)$ gauge sector is added to the SM and the gauge symmetry is Higgsed by a field in the symmetric representation of $SU(N)$.
The Higgs field carries a PQ charge and couples to Weyl fermions transforming under $SU(N)\times SU(3)_c$ as
$
Q:(N,\mathbf{3}), ~ \tilde Q:(N,\bar{\mathbf{3}}),  ~ 
L:(\bar N,\mathbf{1}),  ~  \tilde L:(\bar N,\mathbf{1}).
$
The resulting color anomaly gives a naive domain wall number $N_1=N$.

Reference~\cite{Lu:2023ayc} 
points out that vacuum identification is complete only for odd $N$. From our perspective,  
this reflects the fact that the one-form center symmetry of $SU(N)$ is fully broken by the Higgs only when $N$ is odd. Thus, without computing the domain wall number, one can already exclude even $N$ solely based on the representation of the Higgs (PQ) field. The domain wall number is computed with $N_1=N_2=N$ and $M=2$, leading to $N_{\rm DW}={\rm gcd}(N,M)$. Therefore, we confirm $N_{\rm DW}=1$ only for the odd $N$.

The second setup is the gauged PQ mechanism~\cite{Barr:1992qq,Fukuda:2017ylt}, which introduces a $G=U(1)$ gauge symmetry. Structurally, the model consists of two KSVZ(-like)~\cite{Kim:1979if,Shifman:1979if} sectors, with one linear combination of the two PQ symmetries gauged as $U(1)_{\rm gPQ}$. 
The PQ fields in the ``two'' sectors carry relatively prime charges. Once these fields acquire VEVs, $U(1)_{\rm gPQ}$ is completely Higgsed, 
satisfying the condition for one-form symmetry breaking.  
The domain wall number can then be computed from the master formula, taking into account the possible hierarchy of vacuum expectation values in the PQ sectors. See~\cite{Azatov:2025mep} for a detailed computation.

\section{Physical interpretation of the four-group structure}

We first recall the $M=0$ limit, where the theory factorizes into a BF sector and an axion–three-form sector,
\begin{align}
    S=\frac{1}{2\pi}\int N_2\, B\wedge dA
     +\frac{1}{2\pi}\int N_1\,\phi\, dC .
\end{align}
The topological line and surface operators $e^{i\oint A}$ and $e^{i\oint B}$ create the sources for $dB$ and $dA$, respectively:
\begin{align}
    dB = -\frac{2\pi}{N_2}\,\delta(\gamma),\qquad
    dA = -\frac{2\pi}{N_2}\,\delta(M_2).
\end{align}
This is interpreted that $e^{i\oint A}$ inserts a particle charged under $dB$, while $e^{i\oint B}$ inserts a string magnetically charged under $A$. Their non-trivial correlation between particle and string encodes the Aharonov–Bohm phase.

Similarly, the operators $e^{i \oint C}$ and $e^{i\phi}$ act as sources for $d\phi$ and $dC$,
\begin{align}
    d\phi = -\frac{2\pi}{N_1}\,\delta(M_3), \qquad
    dC = -\frac{2\pi}{N_1}\,\delta(P),
\end{align}
implying that $e^{i\oint C}$ inserts a domain-wall, while $e^{i\phi}$ creates an instanton. Their correlation gives a phase analogous to the AB effect: a domain wall ``sandwiched'' by an instanton–anti-instanton pair acquires a quantized phase.

{\bf $M\neq0$: four-group correlation.}
Turning on $M$ couples the two sectors into a non-trivial four-group structure as discussed before. The relevant topological operators are a line operator $W_1(l)$ and a three-dimensional operator $W_2(D_3)$, both of which now have boundaries.
From the previous interpretation, $W_1(l)$ corresponds to an instanton–anti-instanton pair connected by particles of charge $P N_1/K$, while $W_2(D_3)$ corresponds to a string attached by domain walls of number $P N_2/K$.
If $N_1/ K =1$ 
and $M=1$,
 the object $U_1(V)$ can split into two $W_2$ by pair-creating $e^{i\oint B}$.
This is an interpretation 
of the Lazarides--Shafi mechanism from the 
viewpoint of generalized symmetries.

The mixed correlator $\langle W_1 W_2\rangle$ characterizes the four-group extension. Viewed from the instanton sector, the phase arises from linking between instantons and domain walls; equivalently, from the perspective of probe particles, it appears as the linking between the particle worldline and the string worldsheet. This is a physical manifestation of the non-trivial four-group in our system even without higher-form global symmetries.

\section{Conclusion}\label{sec:conclusion}
We have presented a TQFT action that encodes how the axion vacuum is identified via the center symmetry of a gauge theory.  
Our analysis yields a master formula for systematically computing the domain wall number from the shift of the $\theta$ term and establishes universal conditions on higher-form global symmetries for a successful mechanism.  
In particular, some breaking of the global one-form center symmetry is required to fully identify the axion vacua.  
If this condition is not satisfied, the theory suffers from the domain wall problem.  
While a standard implementation of the mechanism requires the absence of higher-form symmetries, our framework reveals that nontrivial correlations involving the topological operators persist, placing the theory in an SPT phase associated with the four-group structure.
We hope that our results will provide a useful guide for constructing post-inflationary axion models that address both the domain wall problem and axion quality problem through gauge symmetries.

\section*{Acknowledgments}
We are especially grateful to Yichul Choi, Sungwoo Hong, and Yuya Tanizaki for inspiring discussions during the workshop ``Symmetries in Quantum Field Theory and Particle Physics 2025.''
M.S.~also would like to thank Yuta Hamada, Juven Wang, Ling-Xiao Xu, Markus Dierigl for helpful discussions. 
M.S.~is supported by the MUR projects 2017L5W2PT.
M.S. also acknowledges the European Union - NextGenerationEU, in the framework of the PRIN Project “Charting unexplored avenues in Dark Matter” (20224JR28W).
R.Y.~is supported by JSPS KAKENHI Grants No. JP25K17394.

\appendix

\section{Higher-form symmetries for $M\neq 0$}
\label{app:high_sym}
Let us identify genuine (i.e. gauge invariant) point, line, surface, and volume operators for $M\in\mathbb{Z},~M\neq 0$.
The genuine line, volume operators are
\begin{align}
U_1= e^{i\oint C}\ ,~U_2=e^{i\oint A}\ ,
\end{align}

A point operator necessarily comes with an attached line,  
\begin{align}
  W_1(L_1) &=  e^{i\phi(P)} e^{i\frac{PN_2}{K}\int_{L_1}A}e^{-i\phi(P')}\ .
\end{align}
and is therefore not genuine. 
A genuine point operator can be constructed by taking
\begin{align}
    W_1^K\ ,
\end{align}
because 
\begin{align}
  \langle  W_1(l)^K (W_1(l')^{K})^{\dagger}\rangle=1\ ,
\end{align}
where $l'$ is an open line ending with the same points.
This equality follows from the equation of motion for $B$.
There is no genuine operator corresponding to  $W_1^n$ for any integer $n$ with $K>n>0$. 
Concretely, we seek the minimal $n$ satisfying
\begin{align}
    \frac{P}{K} n \in \mathbb{Z}\ .
\end{align}
Using ${\rm gcd}(P,K)=1$,%
\footnote{${\rm gcd}(P,K)=1$: The common integer factor among $M$ and ${\rm gcd}(N_1,N_2)$ is $d={\rm gcd}(N_1,N_2,M)$.
This indicates ${\rm gcd}(P,K)=1$ due to $P=M/d$ and $K={\rm gcd}(N_1,N_2)/d$.}
the minimal $n$ is
\begin{align}
    n=\frac{{\rm gcd}(N_1,N_2)}{{\rm gcd}(N_1,N_2,M)}=K\ .
\end{align}

Similarly, there is a non-genuine surface operator,
\begin{align}
    W_2(D_3)&=e^{i\oint B} e^{-i\frac{P N_1}{K}\int_{D_3} C}\ ,
\end{align}
and the corresponding genuine surface operator is 
\begin{align}
   W_2(D_3)^K 
\end{align}
which satisfies
\begin{align}
   \langle  W_2(D_3)^K ( W_2(D_3')^K )^\dagger\rangle=1\ .
\end{align}

Therefore, the genuine point, line, surface, and volume operators are
\begin{align}
    W_1^K\sim e^{i K\phi},~
    U_2=e^{i\oint A},~
    W_2^K\sim  e^{i K\oint B},~
    U_1=e^{i\oint C}\ .
\end{align}
The correlators 
of $\langle e^{i\oint C}  e^{i K \phi}\rangle$ and $\langle e^{i\oint A}  e^{i K \oint B}\rangle$ 
indicate the symmetries $ \mathbb{Z}_{N_1/K}^{(0)}\times \mathbb{Z}_{N_1/K}^{(3)}$ and $\mathbb{Z}_{N_2/K}^{(1)}\times \mathbb{Z}_{N_2/K}^{(2)}$, respectively.

In summary, the theory has the following global higher-form symmetries,
\begin{align}
    \mathbb{Z}_{N_1/K}^{(0)}\times \mathbb{Z}_{N_2/K}^{(1)}\times \mathbb{Z}_{N_2/K}^{(2)}\times \mathbb{Z}_{N_1/K}^{(3)}\ .
\end{align}

\section{Mathematical structure of four-group}
\label{app:four_group}
From Eqs.~\eqref{eq:U1}--\eqref{eq:w2} we can identify the mathematical structure of the four-group~\cite{Arvasi:2009}
(see also a recent formulation~\cite{fukuda20253crossedmodulesquasicategoriesmoore} and applications in physics~\cite{Hidaka:2021mml,Hidaka:2021kkf}).
This is a so-called strict four-group given by four of groups and 
maps between them, 
\begin{equation}
 G^{(3)}
\overset{\der_3}{\rightarrow} 
 G^{(2)}
\overset{\der_2}{\rightarrow}
 G^{(1)}
\overset{\der_1}{\rightarrow}
 G^{(0)},
\end{equation}
where the groups $G^{(3)}$,..., and $ G^{(0)}$
corresponds to the groups the topological objects
$W_1$, $U_2$, $W_2$, and 
$U_1$, respectively.
The maps $\der_{1,2,3}$ 
specify the correspondences between the parameters of the objects which have boundaries.
In our case, the four-group structure is given by 
\begin{equation}
 \bb{Z}^{(3)}_{N_1}
\overset{\der_3}{\rightarrow} 
\bb{Z}^{(2)}_{N_2}
\overset{\der_2}{\rightarrow}
\bb{Z}^{(1)}_{N_2}
\overset{\der_1}{\rightarrow}
 \bb{Z}^{(0)}_{N_1}
\end{equation}
with maps
\begin{align}
 \der_3 (e^{\fr{2\pi i n_3}{N_1}}) 
&= 
e^{\fr{N_1 P}{K} \cdot \fr{2\pi i n_3}{N_1}}
= 
e^{\fr{N_2 P}{K} \cdot \fr{2\pi i n_3}{N_2}} \in \bb{Z}_{N_2}^{(2)},
\notag
\\ 
 \der_2 (e^{\fr{2\pi i n_2}{N_2}}) 
&= 
1 \in \bb{Z}_{N_1}
\notag
\\ 
 \der_1 (e^{\fr{2\pi i n_1}{N_2}}) 
&= 
e^{ - \fr{N_2 P}{K} \cdot \fr{2\pi i n_1}{N_2}}
= 
e^{ - \fr{N_1 P}{K} \cdot \fr{2\pi i n_1}{N_1}} \in \bb{Z}_{N_1}^{(0)}.
\end{align}
By the maps, we can classify which are genuine symmetry generators of global symmetries.
The genuine symmetry generators are neither nontrivial boundaries nor interiors of some other objects.
The genuine 3-form symmetry generators are parameterized by ${\rm Ker}\, \der_3 = \bb{Z}_{N_1/ K} $ meaning that the image of 
${\rm Ker}\, \der_3$ is trivial. 
The genuine 2-form symmetry 
generators correspond to 
${\rm Ker} \, \der_2 /  \im \, \der_3 = \bb{Z}_{N_2} / \im \, \der_3 $.
Similarly, the 1-form symmetry 
generator are paramterized by 
${\rm Ker} \, \der_1/ \im \, \der_2 
= \bb{Z}_{N_2/ K}$, 
and the 0-form symmetry generators 
are given by 
$\bb{Z}_{N_1} / \im \, \der_1 $.

\bibliography{reference.bib}

\end{document}